\documentstyle[aps,prl]{revtex} 
\draft 
 
\def\lsim{\mathrel{\raise.3ex\hbox{$<$\kern-.75em\lower1ex\hbox{$\sim$}}}} 
\def\gsim{\mathrel{\raise.3ex\hbox{$>$\kern-.75em\lower1ex\hbox{$\sim$}}}}

\newcommand{\simgt}{\lower.5ex\hbox{$\; \buildrel > \over \sim \;$}}
\newcommand{\simlt}{\lower.5ex\hbox{$\; \buildrel < \over \sim \;$}}
\def\change#1{{ #1}}

\begin{document} 
 
\twocolumn[\hsize\textwidth\columnwidth\hsize\csname
@twocolumnfalse\endcsname
 
\title {Possible Evidence for MeV Dark Matter In Dwarf Spheroidals} 
\author{Dan Hooper$^1$, Francesc Ferrer$^{1,2}$, C\'eline Boehm$^1$,
Joseph Silk$^{1,3}$,  Jacques Paul$^4$, N.~Wyn Evans$^5$ and Michel Casse$^{3,4}$} 
\address{
$^1$Astrophysics Department, University of Oxford, Oxford, England UK;
$^2$Theoretical Physics, University of Oxford, Oxford, England UK;
$^3$Institut d'Astrophysique de Paris;
$^4$CEA-Saclay, DSM/DAPNIA/Service d'Astrophysique, F-91191 Gif-sur-Yvette, 
France;
$^5$Institute of Astronomy, University of Cambridge, Cambridge, England UK}
\date{\today} 
 
\maketitle 
 
\begin{abstract}
The observed 511 keV emission from
the Galactic bulge could be due to very light (MeV)
annihilating dark matter particles. To distinguish this hypothesis from conventional astrophysical sources, we study dwarf spheroidals in the region observed by INTEGRAL/SPI such as Sagittarius. As these galaxies have comparatively few stars, the prospects for 511 keV emission from
standard astrophysical scenarios are minimal. The dwarf spheroidals
do, however, contain copious amounts of dark matter. The
observation of 511 keV
emission from Sagittarius should be a ``smoking
gun'' for MeV dark matter.
\end{abstract}

]

\section{Introduction}

Although particle dark matter is generally thought to be in the 10
GeV-1 TeV mass range \cite{lightdarkmatter}, it has been shown that a
1-100 MeV candidate is, in fact, possible \cite{boehm}. Recent
observations of a bright 511 keV $\gamma$-ray line from the Galactic
bulge may be the first experimental evidence for light (1-100 MeV)
annihilating dark matter particles \cite{mev}.

Such particles could annihilate throughout the Galactic bulge
\change{and inner} halo into positrons (and electrons) which, after
losing energy, annihilate into 511 keV gamma-rays.  The observations
of 511 keV emission from the Galactic bulge, made by INTEGRAL
(INTErnational Gamma-Ray Astrophysics Laboratory) \cite{511}, and
previously by CGRO (Compton Gamma Ray Observatory) \cite{previous},
could possibly be explained by a wide variety of astrophysical
scenarios. Proposed sources include neutron stars or black holes
\cite{compact}, radioactive nuclei from supernovae, novae, red giants
or Wolf-Rayet stars \cite{stars}, cosmic ray interactions with the
interstellar medium \cite{ism}, pulsars \cite{pulsars} and stellar
flares.

A popular class of possible sources is type Ia supernovae. The
frequency of such events required to produce a sufficient number of
positrons is $\sim .6$ per century (assuming an escape fraction of 4\%
\cite{late}), however, well above the prediction of current models
(0.03 per century within a factor of 3)
\cite{casse,matteucci,matteucci2}. Alternatively, massive Wolf-Rayet
stars (hypernovae) of the SN2003dh type \cite{sn2003}, exploding in the
Galactic Center are possible candidates \cite{casse}, but their rate
is unknown. Also, even if a very large flux of positrons were injected
into the galactic center, it is not likely that the whole Galactic
bulge could be filled, even if a bipolar galactic wind is produced by
star bursts \cite{starbursts}. If a ``galactic positron fountain''
were to exist \cite{fountain}, the annihilation rate at high altitude
is too low, due to the small density of the wind, to explain the
extension of the 511 keV source \cite{pohl}.

Despite these arguments, it is difficult to be confident that none of
these more standard astrophysical explanations are responsible for the
observed 511 keV line from the Galactic bulge. To more strongly
motivate the light dark matter annihilation scenario, further evidence
is needed.

Dwarf spheroidal galaxies are environments in which high densities of
dark matter are known to be present. Thus, large dark matter
annihilation rates and related gamma-ray fluxes are predicted from
these regions \cite{silk1}. \change{Unlike the Galactic Center, the
dwarf spheroidals are dark matter dominated and do not contain
substantial amounts of gas or stars.} Therefore, observation of bright
511 keV emission from one or more dwarf spheroidals would provide
strong evidence for light annihilating dark matter.

In this paper, we consider the prospects for the observation of 511
keV gamma-ray emission from the two closest dwarf spheroidals
galaxies, Sagittarius and Draco. We find that the flux predicted from
Sagittarius may be above the sensitivity of INTEGRAL/SPI. Therefore,
analysis of the (existing) INTEGRAL/SPI data from this region will
yield a positive signal if light dark matter particles are responsible
for the observed 511 keV flux from the Galactic bulge.

\section{Halo Models}

We parameterize spherical cusped halo profiles \cite{cusp} by
\cite{hernquist}:
\begin{equation}
\rho(r) = \frac{A}{(r/a)^{\gamma}  
[1+(r/a)^{\alpha}]^{(\beta-\gamma)/\alpha}} 
\end{equation}
where $\alpha$, $\beta$ and $\gamma$ are unitless parameters, $a$ is
the distance from the center of \change{the dwarf spheroidal} at which
the power law breaks and $A$ is a normalization constant.

Alternatively, spherical \change{cored} halo models can be
parameterized by \cite{core}
\begin{equation}
\rho(r) = \frac{v_a^2 a^{\alpha} \, 3 r^2+a^2 (1-\alpha)}{4 \pi G (r^2 + a^2)^{2+\alpha/2}} 
\end{equation}
where $\alpha$ is a unitless parameter, $a$ is the core radius and
$v_a$ is the velocity scale. Within the radius $a$, the halo has a
nearly constant density core.

The velocity dispersion of the dwarf spheroidals is largely controlled
by the dark matter density distribution and only weakly affected by
tidal forces (even in the case of the disrupting
Sagittarius)~\cite{tidalpapers}. So, the observational data can be
used to constrain the free parameters in the dark halo
profiles~\cite{ferrer}.  For Draco, the behaviour of the velocity
dispersion with distance is known \cite{kleyna}. Retaining $\gamma$
(for the cusped models) and $\alpha$ (for the cored models) as
arbitrary, the remaining parameters are fixed by requiring that the
velocity dispersion profile of Draco be reproduced.  For Sagittarius,
the velocity data are less complete and the morphological structure
more complicated. However, the central line of sight velocity
dispersion of Sagittarius is similar to that of Draco. Therefore, we
assume that the shape of the halo of Sagittarius is similar to that of
Draco \cite{ferrer}. This simple assumption may be questionable
because tidal disruption has probably distended Sagittarius' dark
matter halo~\cite{jiang} and is relaxed below.

The effect of the dark matter distribution on the annihilation rate
can be described by a single quantity:
\begin{equation}
\overline{J}(\Delta \Omega) \times \Delta \Omega =  \int_{\Delta \Omega} J(\Psi) d \Omega 
\end{equation}
where $\Delta \Omega$ is the solid angle observed, $\Psi$ is the angle
from the center of the halo and $J(\Psi)$ is given by
 \begin{equation} J(\Psi) = \bigg(\frac{1}{0.3 \,
 \rm{GeV/cm}^3}\bigg)^2 \frac{1}{8.5 \, \rm{kpc}} \int_{\rm{los}}
 \rho(r)^2 ds
\end{equation}
where $\rho(r)$ is the dark matter density at a distance $r$ from the
\change{dwarf spheroidal's center} and the integral is performed over
the line of sight of the observation. The rate of annihilations in an
angular region is proportional to $\overline{J}(\Delta \Omega) \times
\Delta \Omega$ and is otherwise independent of the properties of the
halo.

Table I shows the values of $\overline{J}(\Delta \Omega) \times \Delta
\Omega$ for Sagittarius and Draco calculated for several choices of
halo profile, using $\Delta \Omega=$ 0.0038, consistent with the
$2^{\circ}$ angular resolution of SPI. These quantities were computed
following Ref.~\cite{ferrer}. Although the details of these
calculations is beyond the scope of this letter, rough comparisons of
these values can be estimated with simple scaling
relationships. Comparing the fluxes from a dwarf spheroidal and from
the galactic bulge, we estimate
\begin{equation}
\frac{\Phi_{\rm{ds}}}{\Phi_{\rm{gb}}}=\frac{\overline{J}(\Delta \Omega)_{\rm{ds}}  \Delta \Omega_{\rm{ds}}}{\overline{J}(\Delta \Omega)_{\rm{gb}}  \Delta \Omega_{\rm{gb}}} \sim \bigg(\frac{M_{\rm{ds}}}{M_{\rm{gb}}}\bigg)^2  \bigg(\frac{r_{\rm{gb}}}{r_{\rm{ds}}}\bigg)^3 \bigg(\frac{d_{\rm{gb}}}{d_{\rm{ds}}}\bigg)^2,
\end{equation}  
where $M$'s are the masses within a radius $r$, $d$'s are the
distances from Earth and ds and gb denote a dwarf spheroidal and the
galactic bulge, respectively. In this estimate, we have assumed a
fairly flat density within a radius $r$. Using the quantities
$r_{\rm{gb}}\sim \rm{kpc}$, $r_{\rm{ds}}\sim 0.25\,\rm{kpc}$,
$M_{\rm{gb}}(r < \rm{kpc})\sim 10^9 M_{\odot}$, $M_{\rm{ds}}(r < 0.25
\,\rm{kpc})\sim 10^8 M_{\odot}$, $d_{\rm{gb}} \sim 8.5 \, \rm{kpc}$
and $d_{\rm{ds}} \sim 25 \, \rm{kpc}$, we very roughly estimate
$\Phi_{\rm{ds}} \sim 0.1 \, \Phi_{\rm{gb}}$.

The numbers in Table I are reasonably certain for Draco, but can
plausibly be either an order of magnitude larger or smaller for
Sagittarius. Tidal disruption is likely
to have distended the Sagittarius dark halo by a factor of
$\sim 10$. For the same light profile, this causes the values for
Sagittarius in Table I to be increased by a factor of $\sim 30$.

\section{Annihilation and Positron Propagation}

If dark matter particles of $\sim$1-100 MeV mass annihilate into
electron-positron pairs, the resulting positrons then travel,
gradually slowing by ionisation losses. This energy loss rate is
approximately given by \cite{longair} 
\begin{equation}
\label{eq:enlossrate}
\frac{dE}{dt} \sim 2 \times 10^{-12} \bigg(\frac{N_H}{10^2 \rm{m}^{-3}}\bigg) (\ln \Gamma +6.6) \, \rm{eV/s}.
\end{equation}
\noindent
where $\Gamma$ is the positron's Lorentz factor and $N_H$ is the
number density of target atoms. In the Galactic bulge, where we
estimate $N_H \sim 10^5 \rm{m}^{-3}$, this rate can yield stopping
distances of $\sim 10^{24}$ and $\sim 10^{26}$ cm for positrons of MeV
and 100 MeV energy, respectively. No gas has ever been
detected in any of the Local Group dwarf spheroidals. So, the stopping
distance is likely to be significantly longer.
\begin{table}[th]
\begin{tabular}{|c|c|c|} 
Halo Profile & $\overline{J}(\Delta \Omega) \times \Delta \Omega$ Sagittarius & $\overline{J}(\Delta \Omega) \times \Delta \Omega$ Draco  \\ \hline \hline
$\gamma$=1 (NFW) & 0.063  & 0.0057  \\ \hline
$\gamma$=0.8     & 0.063  & 0.0056 \\ \hline
$\gamma$=0.6     & 0.062  & 0.0056 \\ \hline
$\gamma$=0.4     & 0.056  & 0.0050\\ \hline
$\gamma$=0.2     & 0.054  & 0.0049\\ \hline
$\alpha$=0.2     & 0.029 & 0.0026 \\ \hline
$\alpha$=0       & 0.031 & 0.0029 \\ \hline
$\alpha$=-0.2    & 0.034 & 0.0035 \\ 
\end{tabular}
\caption{Values of $\overline{J}(\Delta \Omega) \times \Delta \Omega$
for Sagittarius and Draco calculated using a variety of profiles and
$\Delta \Omega=$ 0.0038, consistent with the $2^{\circ}$ angular
resolution of SPI.}
\end{table}
Although no magnetic fields have been measured in Sagittarius or
Draco, low surface brightness galaxies, which are somewhat similar,
indicate that fields of 2-4 microgauss may be expected
\cite{bdraco}. For microgauss scale magnetic fields, a positron's
Larmour radius is on the order of $10^{11}$ or $10^9$ cm for energies
of 100 MeV and 1 MeV, respectively. Considering a simple random walk,
the positrons are roughly confined to a distance of $\sim
\sqrt{R_{\rm{stop}} \times R_{\rm{Larm}}}$.  Even if we conservatively
estimate magnetic fields with 0.01 microgauss strength and $10^2$
atoms per cubic meter, we find that positrons are stopped within 100
parsecs or less of their generation, a distance much smaller than
could be resolved, given the angular resolution of SPI ($\sim
2^{\circ}$).

The energy loss rate~(\ref{eq:enlossrate}) leads to a thermalization time for
positrons, as a function of energy, of
\begin{equation}
t(E) \sim 10^9 \, \rm{yr} \,\bigg(\frac{E}{\rm{MeV}}\bigg)   \bigg(\frac{10^2 \, \rm{m}^{-3}}{N_e}\bigg) \bigg(\frac{0.5}{f_g}\bigg),
\end{equation}
where $N_e$ is the electron density and $f_g$ is the dust
fraction. Multiplying this by the speed of light, we see that a
positron's stopping distance is typically shorter than its mean free
path. Annihilations are, therefore,
expected to occur primarily for thermalized positrons, producing a 511
keV line. If the dark matter particles are heavier than 50 MeV, or so,
the annihilation time for the positrons produced may exceed the age of
the dwarf spheroid and equilibrium may not be reached, diminishing the
511 keV emission, but probably by less than a factor of 2, or so.

If the electron temperatures are too low, positronium formation may
dominate, resulting in a narrow line (25\% of the time) or 3-photon
continuum (75\% of the time), depending on the spin state of the
positronium. Although positronium formation dominates in the galactic
bulge, as OSSE and INTEGRAL data suggests \cite{milne,strong}, this
may not be the case in a dwarf spheriodal. It is likely, for example,
that the Draco dwarf galaxy is pervaded by diffuse galactic halo gas
at $T\sim 10^6\rm K$ and $N_e \sim 10^{2}(10^6\rm K/T)\, \rm m^{-3},$
as inferred from FUSE observations of high velocity OVI absorption
\cite{sembach}. At this temperature, direct annihilations are
important. If dust is present at even half of the standard
interstellar gas-to-grain ratio, the annihilation line remains narrow
(less than about 2 keV for a grain fraction greater than one tenth of
the local interstellar medium value) \cite{vonballmoos}. Hence the 511
keV line flux should be approximately 4 times greater than in the case
of annihilation through positronium formation, as in the case of the
galactic bulge where the dominant component of the diffuse
interstellar gas is assumed to be at $T\sim 10^4\rm K$.

Given that each annihilating pair of dark matter particles form a
single positron which eventually annihilates producing two 511 keV
gamma-rays, the flux of this gamma-ray line is given by
\begin{equation}
\Phi \cong 5.6\times 2 \times P \, \bigg(\frac{\sigma v}{\rm{pb}}\bigg)
\bigg(\frac{1\,\rm{MeV}}{m_{\rm dm}}\bigg)^2 \overline{J}(\Delta \Omega) \times \Delta \Omega \,\, \rm{cm}^{-2} \rm{s}^{-1},
\end{equation}
where $m_{\rm dm}$ is the mass of the dark matter particle, $\sigma v$
is the annihilation cross section multiplied by the relative velocity
in units of $c$. The quantity $P$ is equal to
0.25 for the galactic bulge (positronium formation) and 1 for dwarf
spheroidals (direct annihilation).

In Ref.~\cite{mev}, it was shown that to explain the angular
distribution of events, as observed by INTEGRAL/SPI, the Galactic halo
is best fit to a mildly cusped profile ($\gamma \sim 0.6$) in the
inner kiloparsecs. The full width, half maximum of the observed
INTEGRAL/SPI signal is $9.0^{+9}_{-3}$ degrees, with 2-$\sigma$
confidence intervals \cite{511}. This corresponds to a value of
$\overline{J}(\Delta \Omega) = 37.6^{+42.1}_{-18.2}$ for $\Delta
\Omega \simeq 0.02$, the angular extent of INTEGRAL's
detection. Combining this with the previous equation, and using the
flux of $9.9 \times 10^{-4}\,\rm{ph}\,\rm{cm}^{-2} \,\rm{s}^{-1}$, as
seen by INTEGRAL, we see that
\begin{equation}
\bigg(\frac{\sigma v}{\rm{pb}}\bigg) \bigg(\frac{1\,\rm{MeV}}{m_{\rm dm}}\bigg)^2 \simeq 4.8^{+4.4}_{-2.4} \times 10^{-4},
\end{equation}
again with 2-$\sigma$ confidence intervals.  We can now combine this
result with values of $\overline{J}(\Delta \Omega) \times \Delta
\Omega$ for specific dwarf \change{spheroidals} to calculate the flux
predicted from such sources.

Considering the range of values for $\overline{J}(\Delta \Omega)
\times \Delta \Omega$ for Sagittarius shown in table I (0.029 to
0.063, for $\Delta \Omega =0.0038$ sr), we can estimate the flux
of 511 keV emission from this region:
\begin{equation}
\Phi \simeq 3.4^{+3.1}_{-1.7} \times 10^{-4} \,\,\rm{to}\,\, 1.6^{+1.5}_{-0.8} \times 10^{-4} \,\, \rm{cm}^{-2} \rm{s}^{-1}.
\end{equation}
The flux from Draco is
approximately a factor of ten smaller. Sagittarius is also near the
Galactic plane, and within the region of the sky which has been
extensively surveyed by INTEGRAL/SPI. For a $3-\sigma$ detection after
an exposure of $10^6$ seconds, the sensitivity of this experiment to
511 keV line emission in this region is estimated to be $\simeq 4
\times 10^{-5}\,\, \rm{cm}^{-2} \rm{s}^{-1}$ \cite{sen}, below our
predicted range of fluxes. We can, therefore, conclude that if the
$\sim 10^{-3} \,\, \rm{cm}^{-2} \rm{s}^{-1}$ flux of 511 keV
gamma-rays observed from the Galactic bulge is the result of light
dark matter annihilations, then analysis of INTEGRAL/SPI's data from
the Sagittarius region of the sky will reveal an observable signal of
511 keV emission. If no such signal is observed, we should consider
other sources for the observed emission from the bulge.

Of course, other potential sources may exist. For example, M31 may
produce a signal similar to that observed from the Galactic bulge. As
M31 is a factor of $\sim100$ further away than the Galactic Center,
however, we expect fluxes $\sim500$ times smaller than from
Sagittarius.

\section{Conclusions}

If light (1-100 MeV) dark matter particles, annihilating to
electron-positron pairs, are responsible for the observed 511 keV
gamma-ray emission from the Galactic bulge, then we expect that
potentially observable fluxes of 511 keV emission would also be
produced in other regions with high dark matter density,
particularly the nearby dwarf spheroidals such as Sagittarius
and Draco. Furthermore, alternative explanations of the
Galactic bulge emission involve exotic stellar objects (hypernovae,
etc.), which are minimal in the directions towards the dwarf
spheroidals. Thus observation of 511 keV emission from such an object
would provide a ``smoking gun'' for annihilating light dark matter
scenarios.

We find that if the observed 511 keV emission from the Galactic bulge
is the product of light annihilating dark matter, then the Sagittarius
dwarf galaxy may provide a 511 keV gamma-ray flux of $\Phi \simeq
3.4^{+3.1}_{-1.7} \times 10^{-4} \,\,\rm{to}\,\, 1.6^{+1.5}_{-0.8}
\times 10^{-4} \,\, \rm{cm}^{-2} \rm{s}^{-1}$. If the dark halo of the
Sagittarius has been distended by tidal forces, then these numbers
could be larger by a factor of $\sim 30$.  Such a flux is above the
sensitivity of INTEGRAL/SPI. If such a signal is seen upon analysis of
the (existing) INTEGRAL/SPI data, it would favor the existence of light scalar dark matter. The absence of such a signal would suggest
that the 511 keV emission observed from the galactic bulge is not
likely to be related to particle dark matter annihilations.

Very recently, a new candidate dwarf spheroidal in the direction of
Canis Major has been suggested. Its
mass and tidal radius have been estimated to be similar to that of the
Sagittarius dwarf ($5 \times 10^8 \, M_{\odot}$ and 2.5 kpc)
\cite{canis}, however, it is
considerably closer and may produce a flux an order of magnitude larger than Sagittarius.

{\it Acknowledgements}: We would like to thank Subir Sarkar for useful
contributions. D.~Hooper and F.~Ferrer are supported by the Leverhulme
trust. C.~Boehm is supported by an individual PPARC fellowship.

\end{document}